\documentclass{article}

\usepackage{emulateapj}
\usepackage{apjfonts}

\newenvironment{inlinetable}{%
\def\@captype{table}%
\noindent\begin{minipage}{0.999\linewidth}\begin{center}\footnotesize}
{\end{center}\end{minipage}\smallskip}

\newenvironment{inlinefigure}{%
\def\@captype{figure}%
\noindent\begin{minipage}{0.999\linewidth}\begin{center}}
{\end{center}\end{minipage}\smallskip}

\newlength{\colwidth}
\newcommand{\cm}{\rm cm} 
\newcommand{\kms}{{\rm km}\,{\rm s}^{-1}}

\newcommand{\HI}{\ion{H}{1}} 
\newcommand{\HeII}{\ion{He}{2}}
\newcommand{\CIV}{\ion{C}{4}} 
\newcommand{\OVI}{\ion{O}{6}}

\newcommand{\lya}{Ly$\alpha$} 
\newcommand{\lyb}{Ly$\beta$}

\lefthead{Schaye et al.}
\righthead{The detection of oxygen in the low-density IGM}

\begin{document}

\submitted{Accepted for publication in ApJ Letters}

\title{The detection of oxygen in the low-density intergalactic
medium\altaffilmark{1}} 
\author{Joop~Schaye\altaffilmark{2}, Michael Rauch\altaffilmark{3},
Wallace L. W. Sargent\altaffilmark{4}, and Tae-Sun
Kim\altaffilmark{5}}
\altaffiltext{1}{Part of the data presented herein were obtained at
the W.~M.~Keck Observatory, which is operated as a scientific
partnership among the California Institute of Technology, the
University of California and the National Aeronautics and Space
Administration. The Observatory was made possible by the generous
financial support of the W.~M.~Keck Foundation.}
\altaffiltext{2}{Institute of Astronomy, Madingley Road, Cambridge CB3
0HA, UK; schaye@ast.cam.ac.uk} 
\altaffiltext{3}{Carnegie Observatories, 813 Santa Barbara Street,
Pasadena, CA 91101; mr@ociw.edu} 
\altaffiltext{4}{Astronomy Department, California Institute of
Technology, Pasadena, CA 91125; wws@astro.caltech.edu} 
\altaffiltext{5}{European Southern Observatory,
Karl-Schwarzschild-Str.\ 2, 85748 Garching, Germany; tkim@eso.org} 

\begin{abstract}
The abundances of metals in the intergalactic medium (IGM) can be used
to constrain the amount of star formation at high redshift and the
spectral shape of the ionizing background radiation. For both purposes
it is essential to measure the abundances in regions of low density, away
from local sources of metals and ionizing photons. Here we report the
first detection of \OVI\ in the low-density IGM at high redshift. We
perform a pixel-by-pixel search for \OVI\ absorption in eight high
quality quasar spectra spanning the redshift range $z=2.0$-4.5. At
$2 \la z\la 3$, we clearly detect \OVI\ in the form of a positive correlation
between the \HI\ \lya\ optical depth and the optical depth in the
corresponding \OVI\ pixel, down to $\tau_{\rm HI} \sim 10^{-1}$. This
is an order of magnitude lower in $\tau_{\rm HI}$ than the best \CIV\
measurements can probe and constitutes the first clear detection of
metals in \emph{underdense} gas. The non-detection of \OVI\ at $z > 3$
is consistent with the enhanced photoionization from a hardening of the
UV background below $z\sim 3$ but could also be caused by the high
level of contamination from Ly series lines. 
\end{abstract}


\section{Introduction}
The abundance of metals in the low-density intergalactic medium (IGM)
at high redshift is of interest for at least two reasons. Firstly, the
metallicity can be used to distinguish between different enrichment
mechanisms and to constrain the amount of star formation at high
redshift. Secondly, the relative abundances of ions with different
ionization potentials can be used to derive the spectrum of the
integrated UV background from stars and quasars. For both purposes it
is essential to measure the metal abundances in gas of low density,
away from the influence of local sources of ionizing radiation and
locally produced metals.

Resonant \lya\ absorption by neutral hydrogen along the line of sight
to distant quasars results in a forest of absorption lines bluewards
of the quasar's \lya\ emission line. With the advent of the
High-Resolution Echelle Spectrograph (HIRES)
on the Keck telescope, it became clear that many of the
high column density \lya\ lines ($N_{\rm HI} \ga 10^{14.5
}\,\cm^{-2}$) show associated \CIV\ absorption (e.g., Cowie et
al.~1995). Although increasing the signal-to-noise ratio (S/N) reveals
\CIV\ systems of progressively lower column density, \CIV\ lines
corresponding to low column density \lya\ lines are generally far too
weak to detect (e.g., Ellison et al.\ 2000).

The success of cosmological simulations in reproducing the
observations of the \lya\ forest has convincingly shown that the low
column density forest ($N_{\rm HI} \la 10^{14.5}\,\cm^{-2}$) at high
redshift ($z \ga 2$) arises in a smoothly fluctuating IGM, with
individual lines corresponding to local density maxima of moderate
overdensity ($\rho/\bar{\rho} \la 10$). The interplay between
photoionization heating and adiabatic cooling that is due to the universal
expansion results in a tight temperature-density relation, which is
well approximated by a power-law, $T=T_0(\rho/\bar{\rho})^{\gamma-1}$
(Hui \& Gnedin 1997). The \lya\ optical depth is proportional to the
neutral hydrogen density, which in photoionization equilibrium is
proportional to $\rho^2 T^{-0.76}$. Hence the optical depth depends on
the underlying baryon overdensity, $\tau_{\rm HI} \propto
(\rho/\bar{\rho})^{2.76-0.76\gamma}$. The constant of proportionality
is an increasing function of redshift, at $z\sim 3$, an optical depth
of one corresponds to slightly overdense gas.

Cowie \& Songaila (1998) realized that the tight correlation between
optical depth and gas overdensity makes a pixel-by-pixel analysis of
metal abundances useful. By measuring the median \CIV/\HI\ optical
depth ratio as a function of $\tau_{\rm HI}$, they were able to show
that the IGM is enriched with a roughly constant \CIV/\HI\ ratio down
to $\tau_{\rm HI}\sim 1$ at $z\sim 3$. Ellison et al.~(2000) showed
explicitly that directly detected \CIV\ lines account for only a small
fraction of the total amount of metals in the forest, as inferred from
a pixel analysis.

Although other metal lines have been detected, \CIV\
($\lambda\lambda$1548, 1551) has so far proven the most sensitive
since it is the strongest line redwards of \lya\ ($\lambda$1216),
where there is little contamination from other absorption
lines. However, simulations show that at high redshift, photoionized
\OVI\ ($\lambda\lambda$1032, 1038) should be a much better probe of
the metallicity of the IGM in regions close to the mean
density (Chaffee et al.~1986; Rauch, Haehnelt, \& Steinmetz 1997;
Hellsten et al.~1998). In practice it has proven very difficult to
detect \OVI\ at high redshift because it lies deep in the \lya\ and
\lyb\ ($\lambda 1026$) forest. Since Lu \& Savage (1993) established
the presence of \OVI\ in intervening absorbers by stacking Lyman limit
systems that show \CIV\ absorption, \OVI\ has been detected at high
redshift in several Lyman limit systems (Vogel \& Reimers 1995,
Kirkman \& Tytler 1997; 1999). Dav\'e et al.~(1998) (see also Cowie \&
Songaila 1998) carried out a thorough search for \OVI\ but found only
evidence for \OVI\ associated with \CIV\ absorption in high density
gas.
  
At low redshift ($z \la 1$), \OVI\ absorbers have been shown to be
common (e.g., Bergeron et al.\ 1994), so common in fact that they may
harbor a large fraction of the 
baryons (Burles \& Tytler 1996; Tripp, Savage \& Jenkins 2000). These
low redshift systems probably correspond to hot, shock heated gas in
the potential wells of (groups of) galaxies, which may well be
collisionally ionized.

We have carried out a pixel-by-pixel search for \OVI\ in the spectra
of eight quasars, spanning the redshift range $z\sim 2.0$--4.5. We
clearly detect \OVI\ at $2\la z \la 3$, down to very low \HI\ optical
depths, $\tau_{\rm HI} \sim 10^{-1}$, which is an order of magnitude
lower than the best \CIV\ measurements can probe.

\section{Observations and sample definition}
We analyzed spectra of the eight quasars listed in column 6 of
Table~1. The spectra of Q1101, J2233 and Q1122 were taken during the
Commissioning I and Science Verification observations of the UV-Visual
Echelle Spectrograph (UVES) and have been
released by ESO for public use. They were reduced with the
ESO-maintained MIDAS ECHELLE package (see Kim, D'Odorico, \& Cristiani
2000, in preparation) for details on the data reduction). The other spectra
were obtained with the HIRES spectrograph (Vogt et al.~1994) on the
Keck telescope. The reduction procedures for these quasars can be
found in Barlow \& Sargent~(1997) and Sargent, Barlow \& Rauch (2000, in
preparation). The HIRES and UVES spectra have a nominal velocity
resolution of 6.6 and 6.7~$\kms$ (FWHM) and a pixel size of 0.04 and
0.05~\AA\ respectively. To avoid confusion with the \lyb\ forest, we
only used the \lya\ region between the quasar's \lya\ and \lyb\
emission lines. In addition, spectral regions close to the quasar
(typically $5000~\kms$) were excluded to avoid proximity
effects. Regions in the \lya\ forest thought to be contaminated by
metal lines were excluded.

The mean absorption increases rapidly bluewards of the quasar's \lyb\
emission line becaue of the growing number of absorption lines other
than \lya. For quasars at $z \ga 3$ the increase in the absorption is
dominated by higher order Ly lines, corresponding to \lya\ lines which
fall in between the quasar's \lyb\ and \lya\ emission lines (i.e.\ in
the \lya\ forest). At lower redshift the hydrogen lines have smaller
column densities and metal lines contribute significantly to the mean
absorption. The increase in the mean absorption towards the blue,
raises the threshold for detecting \OVI\ absorption in the \lya\
forest considerably. Moreover, the S/N in the \OVI\ region generally
decreases towards the blue, raising the detection threshold further.
We therefore analyzed the red and blue halves, i.e.\ the high and low
redshift halves, of each \lya\ forest spectrum separately (for those
quasars for which our spectral coverage extends far enough into the
blue). This turned out to be crucial, as we were unable to detect
\OVI\ in any of the low redshift half samples. All the high redshift
half samples are listed in Table~1.

\vspace{0.5cm}
\begin{inlinetable}
\begin{tabular}{lccrrcc}
\multicolumn{7}{c}{TABLE 1}\\
\multicolumn{7}{c}{\sc List of samples}\\
\tablevspace{0.1cm}
\tableline
\tableline
\tablevspace{0.1cm}
Sample & $z_{\min}$ & $z_{\max}$ & (S/N)$_{\rm OVI}$ & (S/N)$_{\rm
HI}$ 
& QSO & $z_{\rm em}$\\
\tablevspace{0.1cm}
\tableline
1101 & 1.97 & 2.10 & 21 & 78 & Q1101$-$264 & 2.14 \\
2233 & 2.07 & 2.18 & 12 & 30 & J2233$-$606 & 2.24 \\
1122 & 2.03 & 2.34 & 27 & 66 & Q1122$-$165 & 2.40 \\
1442 & 2.51 & 2.63 & 8 & 61 & Q1442+293 & 2.67 \\
1107 & 2.71 & 2.95 & 10 & 49 & Q1107+485 & 3.00 \\
1425 & 2.92 & 3.14 & 42 & 113 & Q1425+604 & 3.20 \\ 
1422 & 3.22 & 3.53 & 46 & 105 & Q1422+231 & 3.62 \\
2237 & 4.15 & 4.43 & 31 & 32 & Q2237$-$061 & 4.55 \\ 
\tableline
\end{tabular}
\label{tbl:samples}
\end{inlinetable}

\section{Method}
\label{sec:method}
We search for \OVI\ using a variant of the pixel technique introduced
by Cowie \& Songaila (1998). We measure the optical depth in each
\lya\ forest pixel and in the corresponding \OVI\ pixel. The \HI-\OVI\
pixel pairs are binned according to their \HI\ optical depth and the
median \HI\ and \OVI\ optical depths are computed for each bin. This
technique enables us to probe the \OVI\ abundance down to much lower
densities than is possible by fitting lines. Its main advantage over,
for example, the stacking method is its robustness. Being a median, it
is relatively insensitive to (non-Gaussian) noise, contamination from
non-\OVI\ lines, problems that severely compromise the stacking
method, which is a mean (see Ellison et al.~2000 for a critical
assessment of both the stacking technique and the pixel technique for
the case of \CIV). Its main limitation is that it measures an
\emph{apparent} \OVI\ optical depth, $\tau_{\rm OVI, app}$, which is
in general not the same as the true \OVI\ optical depth because of
contamination.

The effect of contaminating lines is reduced by defining $\tau_{\rm
OVI, app} \equiv \min(\tau_{\rm OVI,\,1032},2\tau_{\rm OVI,\,1038})$,
where we used the fact that $(f\lambda)_{1032} =
2(f\lambda)_{1038}$. Taking the minimum doublet component does of
course not guarantee that $\tau_{\rm OVI, app} = \tau_{\rm OVI}$. For
some pixels both components may be contaminated, which would result in
an overestimate of the \OVI\ optical depth. However, if the
contamination is negligible, then noise will cause us to underestimate
$\tau_{\rm OVI}$. For \CIV, which falls redwards of the \lya\ forest,
noise is the limiting factor and Ellison et al.~(2000) therefore only
used the weaker component if a $3\sigma$ detection was predicted,
based on the signal in the stronger component. For \OVI\ however,
contamination is a much greater problem and we therefore take the the
minimum if the expected signal in the weaker component is greater than
0.5$\sigma$, i.e.\ if $\exp(-0.5\tau_{\rm OVI,\,1032}) <
1-0.5\sigma_{\rm OVI,\,1038}$, where $\sigma_{\rm OVI,\,1038}$ is the
rms amplitude of the noise at the position of \OVI,\,1038 (as given by
the normalized noise array).

When a pixel is close to saturation [$\exp(-\tau_{\rm HI}) <
0.5\sigma$], we use up to 10 higher order Ly lines to determine the
optical depth: $\tau_{{\rm Ly}\alpha}\equiv \min(\tau_{{\rm
Ly}n}f_{{\rm Ly}\alpha}\lambda_{{\rm Ly}\alpha}/f_{{\rm
Ly}n}\lambda_{{\rm Ly}n})$. We limit the effect of noise features by
requiring $0.5\sigma_n < \exp(-\tau_{{\rm Ly}n}) < 1-0.5\sigma_n$
($\sigma_n$ is the rms noise amplitude at the position of Ly$n$).  The
number of higher order lines available is different for each quasar
spectrum and varies also within a single spectrum. However, since the
number of pixels with a given $\tau_{\rm HI}$ decreases rapidly for
$\tau_{\rm HI} \ga 1$, the misbinning of \HI\ due to missing higher
orders affects only the highest $\tau_{\rm HI}$ bins.
 
We use logarithmic $\tau_{\rm HI}$ bins of size 0.35~dex (0.5~dex for
the two lowest redshift quasars) and estimate the error in the medians
by bootstrap resampling\footnote{The spectrum is divided into chunks
of 2\,\AA\ and 250 spectra of length equal to the original spectrum
are generated by picking chunks at random (with replacement, each
chunk can be picked more than once). The $1\sigma$ error in the median
is then the square root of the variance in the median over the
bootstrap realizations.}.

\vspace{1cm}
\section{Results}
\begin{figure*}[t]
\begin{center} 
\resizebox{0.96\textwidth}{!}{\includegraphics{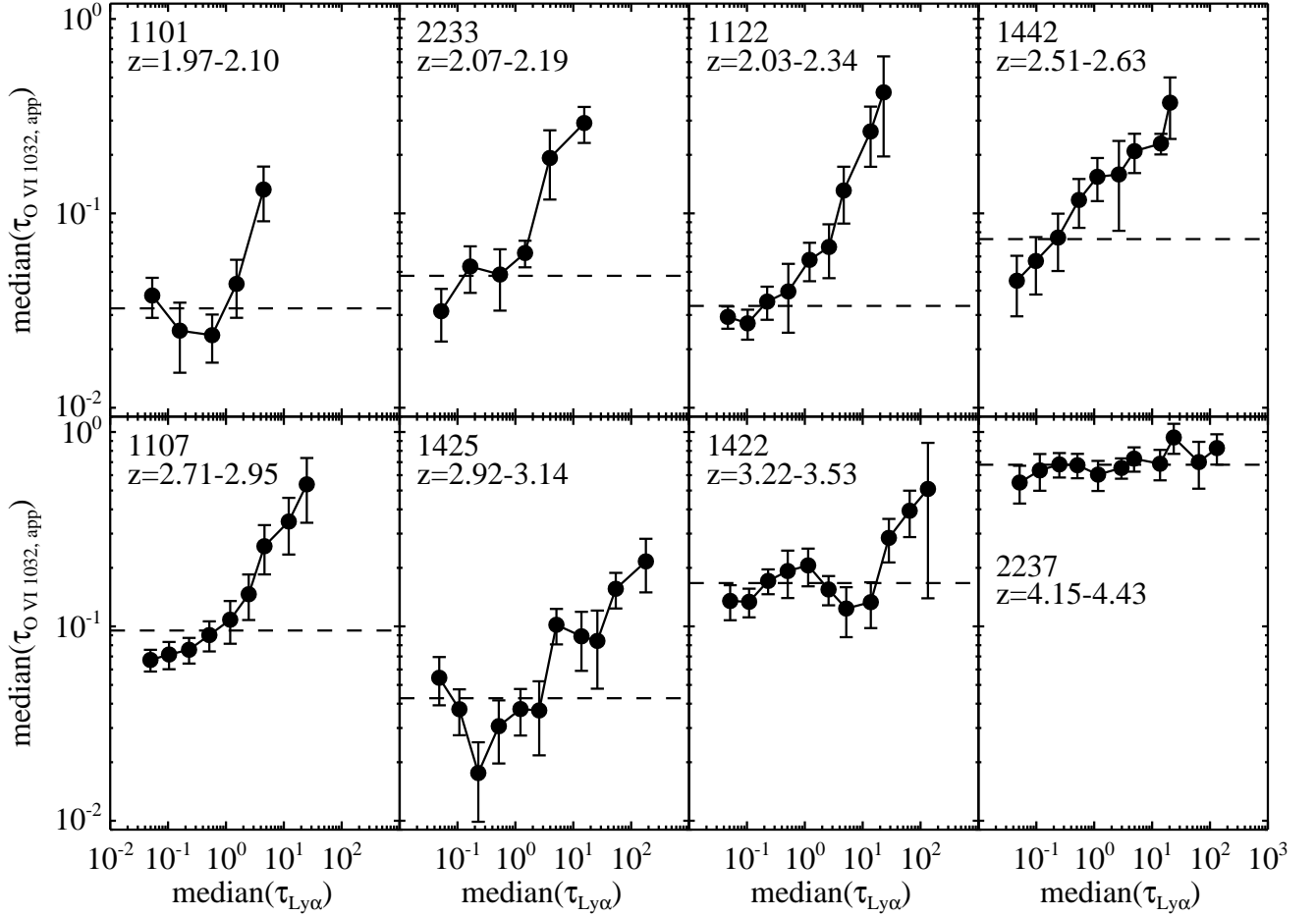}}
\figcaption[schaye.fig1.eps]{Results from the pixel analysis. For all
HI \lya\ pixels within the redshift range indicated in each panel, the
corresponding OVI pixel was determined and the optical depths were
measured.  Pixels corresponding to higher order Ly lines and to the
weaker line of the OVI doublet were used as described in section
\protect\ref{sec:method}. HI-OVI pixel pairs were binned according to
their HI optical depth and the data points indicate the median HI and
OVI optical depth for each bin. The measured, apparent OVI optical
depths include contributions from noise and contaminating non-OVI
lines. The dashed lines show the median, apparent OVI optical depth
for the full set of HI pixels. If the contribution of OVI to the
median absorption is negligible, then the dashed lines are effectively
the OVI detection limits. Vertical error bars are $1\sigma$ errors,
horizontal error bars are smaller than the symbols and are not
shown. For $z\la 3$ $\tau_{\rm OVI, app}$ and $\tau_{\rm HI}$ are
clearly correlated, down to optical depths as low as $\tau_{\rm HI}
\sim 10^{-1}$. A correlation between $\tau_{\rm OVI, app}$ and
$\tau_{\rm HI}$ implies that OVI absorption has been detected in the
Ly$\alpha$ forest.}
\label{fig:uphalf}
\end{center}
\end{figure*}

Figure~1 shows the results from a pixel analysis of the samples listed
in Table~1. The apparent \OVI\ optical depth is clearly correlated
with $\tau_{\rm HI}$ at $z\la 3$ down to $\tau_{\rm HI} \sim 10^{-1}$,
indicating that \OVI\ is detected in the \lya\ forest.  At $z\sim
3.2$--3.5 we detect \OVI\ only at very high \HI\ optical depths,
$\tau_{\rm HI} > 10$, while at $z > 4$ $\tau_{\rm OVI}$ and $\tau_{\rm
HI}$ appear uncorrelated. The transition seems to occur at $z\sim
3.0$--3.5. If contamination and noise, which are independent of
$\tau_{\rm HI}$, dominate over absorption from \OVI, then $\tau_{\rm
OVI, app}$ will flatten off at the detection limit. Ellison et al.\
(2000) analyzed a very high quality spectrum of Q1422+231 ((S/N)$_{\rm
CIV} \sim 200$) and found that $\tau_{\rm CIV}$ flattens off below
$\tau_{\rm HI} \sim 2$. We find that the correlation between
$\tau_{\rm OVI, app}$ and $\tau_{\rm HI}$ continues down to much lower
$\tau_{\rm HI}$, indicating that oxygen has been detected in gas of
very low density contrast ($\rho/\bar{\rho} < 1$).

The horizontal, dashed lines indicate the median $\tau_{\rm OVI, app}$
corresponding to a random \HI\ pixel, i.e.\ the median $\tau_{\rm OVI,
app}$ for the set of all \HI\ pixels. This reference level includes
contributions from \OVI, from contaminating non-\OVI\ lines and from
noise. For $z \ga 3$, the contribution of \OVI\ to this level is
negligible and the level is effectively the detection limit. For $z
\la 3$ on the other hand, \OVI\ accounts for a significant part of the
total absorption. Hence the detection limit is lower than the
reference level and $\tau_{\rm OVI, app}$ drops below the dashed line
at low $\tau_{\rm HI}$.

We stress that because of noise and especially contamination, the
measured, apparent \OVI\ optical depths could differ considerably from
the true values. We tried to estimate the effects of noise and
contamination, which vary systematically across the spectrum, by
analyzing synthetic spectra generated by drawing \HI\ Voigt profiles
at random from observed line lists until the mean absorption in the
\lya\ forest equals the observed value. We then added the
corresponding \OVI\ absorption lines using a fixed \OVI/\HI\ column
density ratio and a fixed $b$-parameter ratio. Finally, higher order
Ly lines and noise were added using the \HI\ line list and the noise
array respectively. We found that the simulations agree reasonably
well with the observations for $N($\OVI$)/N($\HI$) \sim
10^{-2}$--$10^{-1}$, although the results are sensitive to the assumed
ratio of $b$-parameters. Although encouraging, these Monte Carlo
simulations are clearly too simplistic to derive the intrinsic
\OVI/\HI\ ratios. The effects of noise and contamination can only be
modeled accurately by calibrating the method against synthetic spectra
extracted from hydrodynamic simulations. This should be done
separately for each observed spectrum, using synthetic spectra that
resemble the observed spectrum in detail. We leave such a quantitative
analysis for a future paper but note that the effects that prevent us
from deriving the actual \OVI\ abundance, will reduce any intrinsic
correlation between $\tau_{\rm HI}$ and $\tau_{\rm OVI, app}$, making
a detection of such a correlation a robust result.

For a fixed \OVI\ abundance, the value of the reference level (Fig.\ 1,
dashed lines) increases with decreasing S/N and with increasing redshift. The
latter effect is more important and is due to the increase in the
number and strength of contaminating Ly series lines with redshift,
which is a direct consequence of the expansion of the universe. As the
universe expands, the column densities of lines corresponding to
absorbers of a fixed overdensity decrease (the evolution of the
ionizing background and the growth of structure are less
important). This results in an increase of the amount of
contamination, and thus the detection threshold, with redshift. Note
that this effect is opposite to the increase in contamination towards
the blue (i.e.\ low redshift) end of a single quasar spectrum, which
is a consequence of the growing number of interlopers (see section
2). Like the column density, the \HI\ optical depth corresponding to a
fixed overdensity also increases with redshift. This complicates the
comparison of the $\tau_{\rm HI}$ level below which the correlation
with \OVI\ flattens off in samples of different redshifts.  While an
\HI\ optical depth of one corresponds to about the mean density at
$z\sim 3$, it corresponds to an overdensity of a few at $z\sim 2$ and
to slightly underdense gas at $z\sim 4$.

For the quasars Q1107, Q1422 and Q2233 our spectral coverage is
sufficient to analyze the blue half of the forest. We find no
correlation between $\tau_{\rm OVI}$ and $\tau_{\rm HI}$ in any of
these samples. As discussed above, this is a consequence of the large
number of contaminating absorption lines from intervening absorbers
and/or the low S/N of the data and does not imply that there is no
\OVI. In principle, even more \OVI\ could be uncovered by looking at a
smaller part of the spectrum. The contamination in the upper redshift
quarter for example, will be lower than in the upper half. However,
decreasing the number of pixels further would bring the redshift path
into the regime of large-scale structures and the samples would
therefore no longer be representative. This may already be the case
for the two lowest redshift samples, for which the number of pixels
showing \HI\ absorption is much smaller than for the higher redshift
samples. Nevertheless, we have tried smaller samples and found that
the correlation strengthens somewhat for 1422 but not for 2237.

Figure~2 demonstrates that the \OVI\ that we detect at low $\tau_{\rm
HI}$ is not associated with detectable \CIV\ absorption. The solid
line indicates the apparent \OVI\ absorption in 1442. The dashed line
shows the result obtained after all pixels that have nonzero \CIV\
absorption at the $1\,\sigma$ level [$\exp(-\tau_{\rm CIV}) <
1-1\sigma$] have been excluded. The dashed curve only falls below the
solid curve for $\tau_{\rm HI} \ga 2$, indicating that some of the
stronger \OVI\ systems have been excluded. The fact that the two
curves are almost indistinguishable for $\tau_{\rm HI} \la 2$ shows
that \OVI\ is detectable in absorbers for which \CIV\ absorption is
too weak to detect and is thus direct observational proof of the
prediction that \OVI\ is a more sensitive probe of the metallicity in
low-density gas than \CIV.
\begin{inlinefigure}
\centerline{\resizebox{0.96\colwidth}{!}{\includegraphics{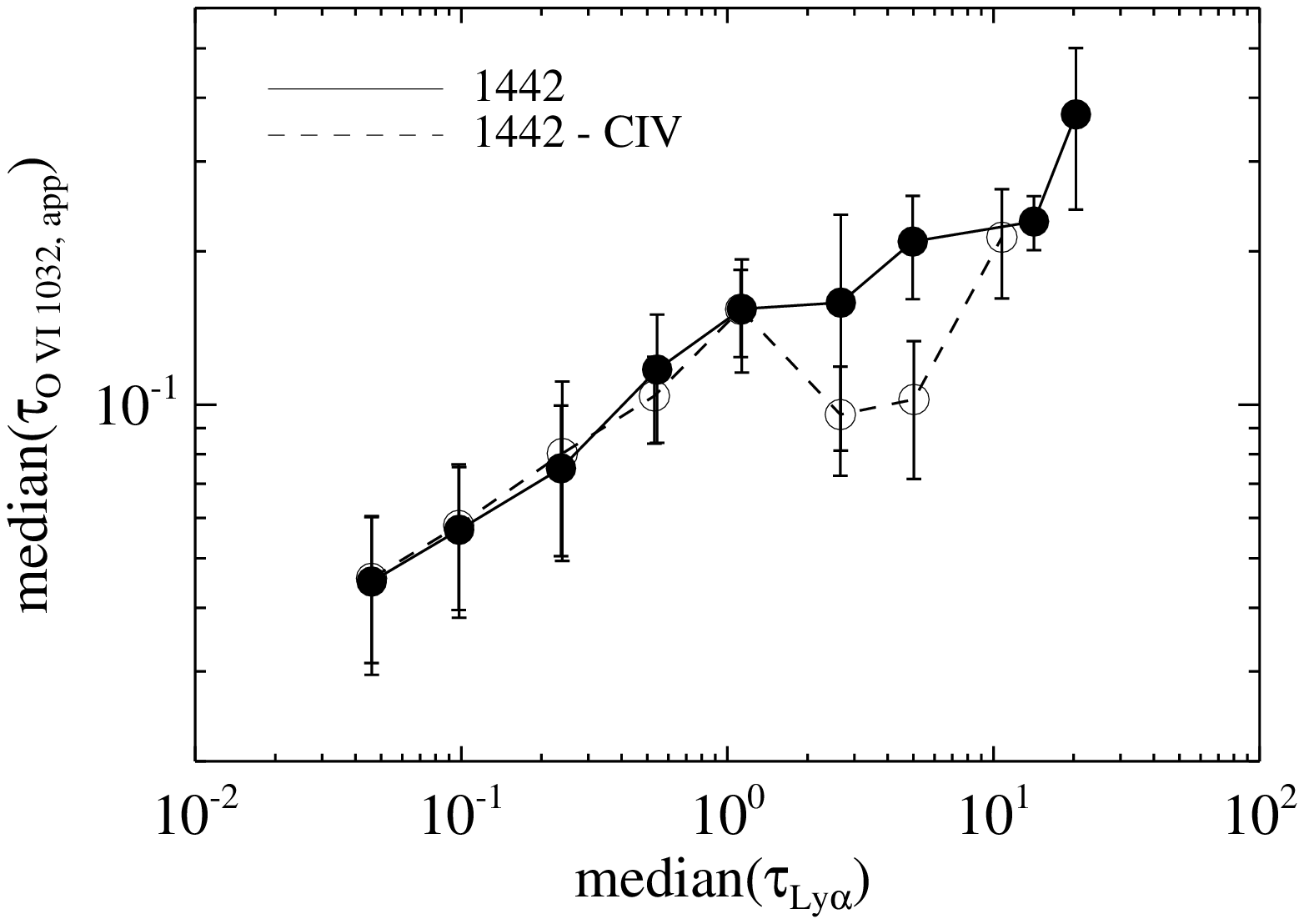}}}
\figcaption[schaye.fig2.eps]{The effect of excluding CIV
absorbers. The solid curve is for sample 1442. The dashed curve shows
the result of the pixel analysis after having excluded all pixels for
which the corresponding CIV absorption is detectable. The curves only
start to differ at $\tau_{\rm HI} \ga 2$, indicating that only the
strong OVI absorbers show significant CIV absorption.}
\label{fig:civ}
\end{inlinefigure}

\section{Discussion}
We have searched for \OVI\ absorption in the spectra of eight quasars
spanning the redshift range $z\sim 2.0$--4.5 and reported the first
detection of \OVI\ in the low-density IGM at high redshift. We
analyzed the spectra pixel by pixel and detected a positive
correlation between the optical depths in \HI\ and in the
corresponding \OVI\ pixels down to $\tau_{\rm HI}\sim 10^{-1}$ at $2
\la z\la 3$. This is an order of magnitude lower in $\tau_{\rm HI}$
than the best \CIV\ measurements can probe and constitutes the first
firm detection of metals in \emph{underdense} gas. We showed that the
\OVI\ signal does not come from systems for which \CIV\ is detectable,
and this confirms the prediction that \OVI\ is a much better tracer of
metals in low-density gas than \CIV.

Although the very strong absorbers may well be collisionally ionized
(e.g., Kirkman \& Tytler~1999), the volume filling factor of this hot
gas is much smaller than that of photoionized gas (e.g., Cen \&
Ostriker 1999). The ubiquitous \OVI\ detected here, arises in gas that
is too dilute for collisional ionization to be effective. Photons of
at least 8.4\,Ryd are needed to create \OVI\ and hence we would not
expect to detect photoionized \OVI\ if the reionization of \HeII, for
which photons of at least 4\,Ryd are required, is not yet
complete. Measurements of the \HeII\ \lya\ opacity (Heap et al.~2000
and references therein) and the temperature of the IGM (Schaye et al.\
2000, Ricotti, Gnedin \& Shull 2000) suggest that the reionization of
helium may have been completed at $z\sim 3$. Although our
non-detection of \OVI\ at $z > 3$ is consistent with enhanced
photoionization from a hardening of the ionizing background below
$z\sim 3$, it could also be caused by the increased level of
contamination from Ly series lines.

Our detection of \OVI\ down to $\tau_{\rm HI}\sim 10^{-1}$ shows that
the IGM is enriched down to much lower overdensities than have been
probed up till now. A comparison with simulations could clarify how
noise and contamination from Ly series lines affect the apparent \OVI\
optical depth, what fraction of the low-density gas is enriched and
what the scatter in the \OVI/\HI\ ratio is, thereby providing strong
constraints on theoretical models for the enrichment of the IGM.

\acknowledgments We are grateful to S.~Ellison, M.~Pettini and
G.~Efstathiou for discussions and a careful reading of the
manuscript. JS thanks the Isaac Newton Trust and PPARC for
support. WLWS acknowledges support from the NSF under grant
AST-9900733.

\end{document}